\documentclass[aps,prd,twocolumn,superscriptaddress,nofootinbib,floatfix,noshowpacs]{revtex4-2}
\pdfoutput=1
\newif\ifcomment
\usepackage{amsmath,amssymb}
\usepackage{color,hyperref,url}
\usepackage{listings}
\usepackage{slashed}
\usepackage[pdftex]{graphicx}
\usepackage{epstopdf}
\usepackage{epsfig}
\usepackage{grffile}
\usepackage{relsize}
\graphicspath{{./img/}}
\usepackage{soul}
\usepackage{xcolor}
\newcommand{\beq}{\begin{equation}}
\newcommand{\eeq}{\end{equation}}
\newcommand{\ba}{\begin{array}}
\newcommand{\ea}{\end{array}}
\newcommand{\bea}{\begin{align}}
\newcommand{\eea}{\end{align}}
\newcommand{\bi}{\begin{itemize}}
\newcommand{\ei}{\end{itemize}}
\newcommand{\ben}{\begin{enumerate}}
\newcommand{\een}{\end{enumerate}}
\newcommand{\bc}{\begin{center}}
\newcommand{\ec}{\end{center}}
\newcommand{\bl}{\begin{flushleft}}
\newcommand{\el}{\end{flushleft}}
\newcommand{\br}{\begin{flushright}}
\newcommand{\er}{\end{flushright}}

\newcommand{\nn}{\nonumber \\}
\newcommand\Eqn[1]{Eq.~(\ref{#1})}  % includes ``Eq.'' in front
\newcommand{\fm}{{\rm fm}}

\newcommand{\eV}{{\rm eV}}

\newcommand{\MeV}{{\rm MeV}}
\newcommand{\GeV}{{\rm GeV}}

\begin{document}
\title{Neutral pion to two-photons transition form factor revisited}

\author{M. Atif Sultan}%
\email{atifsultan.chep@pu.edu.pk}
\affiliation{School of Physics, Nankai University, Tianjin 300071, China}
\affiliation{Centre For High Energy Physics, University of the  Punjab, Lahore (54590), Pakistan}

\author{Jiayin Kang}%
\email{kangjiayin@mail.nankai.edu.cn}
\affiliation{School of Physics, Nankai University, Tianjin 300071, China}

\author{Adnan Bashir}%
\email{adnan.bashir@umich.mx}
\affiliation{Instituto de F\'isica y Matem\'aticas, Universidad Michoacana de San Nicol\'as de Hidalgo, Morelia, Michoac\'an 58040, M\'exico}
\affiliation{Department of Integrated Sciences and Center for Advanced Studies in Physics, Mathematics and Computation, University of Huelva, E-21071 Huelva, Spain}

\author{Lei Chang}%
\email{leichang@nankai.edu.cn}
\affiliation{School of Physics, Nankai University, Tianjin 300071, China}
\date{\today}
%============================================================
\begin{abstract}

Based upon a combined formalism of Schwinger-Dyson and Bethe-Salpeter equations in quantum chromodynamics (QCD), we propose a QCD kindred algebraic model for the dressed quark propagator, for the 
Bethe-Salpeter amplitude of the pion and the electromagnetic quark-photon interaction vertex. We then compute the $\gamma^{*}\pi^0\gamma$ transition form factor $G^{\gamma^{*}\pi^0\gamma}(Q^2)$ for a wide range of photon momentum transfer squared $Q^2$. The quark propagator is expanded out in its perturbative functional form but with dynamically generated dressed quark mass. It has complex conjugate pole singularities in the complex-momentum plane which is motivated by the solution of the quark gap equation with rainbow-ladder truncation of the infinite set of Schwinger-Dyson equations. This complex pole singularity structure of the quark propagator can be associated with a signal of confinement which prevents quarks to become stable asymptotic states. %The Bethe-Salpeter amplitude is expressed in terms of a spectral density function which is fully constrained  by the distribution amplitude of the pion, which in turn is well understood both at the hadronic scale and in asymptotic QCD. 
The Bethe-Salpeter amplitude is expressed without a spectral density function, which encapsulate its low and large momentum behaviour. The QCD evolution of the distribution amplitude is also incorporated into our model through the direct implementation of Efremov-Radyushkin-Brodsky-Lepage evolution equations. We include the effects of the quark anomalous magnetic moment in the description of the quark-photon vertex whose infrared enhancement is known to dictate hadronic properties. Once the QCD kindred model is constructed, we calculate the form factor $G^{\gamma^{*}\pi^0\gamma}(Q^2)$ and find it consistent with direct QCD-based studies as well as most available experimental data. It slightly exceeds the conformal limit for large $Q^2$ which might be attributed to the scaling violations in QCD. The associated interaction radius and neutral pion decay width turn out to be compatible with experimental data.

\end{abstract}

\maketitle
%============================================================
\section{introduction}\label{sec:int}
Quantum chromodynamics (QCD)~\cite{Fritzsch:1973pi}, the theory of strong interactions, is perhaps the most intriguing and challenging piece of the Standard Model of elementary particles. Recall that all the QCD elementary degrees of freedom, quarks and gluons, lie in the fundamental and the adjoint representation, respectively, of the $SU(3)$ color gauge group.  
 Therefore, it is a non-{\em Abelian} gauge theory; it is  renormalizable and it exhibits asymptotic freedom~\cite{Gross:1973id, Politzer:1973fx} in its perturbative tail. This feature allows us to study strong interaction processes using well established Feynman diagrammatic approach at energy scale sufficiently greater than its intrinsic hadronic scale~$\Lambda_{\rm QCD}$. However, in the infrared region, strong coupling is of order one or higher which makes the perturbative approach inapplicable. 
 Therefore, at this energy scale one is forced to resort to intrinsically non-perturbative methods such as lattice simulation of QCD~\cite{wilson1974confinement} and a coupled formalism of Schwinger-Dyson (SDEs) and Bethe-Salpeter equations (BSEs)~\cite{Dyson:1949bp, Schwinger:1951ex}.\par

All hadronic observables may be calculated if we know the Green functions of QCD which satisfy a set of infinite, coupled, nonlinear integral equations, the already mentioned SDEs~\cite{Dyson:1949ha,Schwinger:1951ex}. The mathematical structure of these equations is such that the two-point one-particle irreducible (1PI) Green functions (quark, gluon and ghost propagators) are related to the three-point functions (for example, quark-gluon and triple gluon vertices), which in turn are intrinsically entangled with the four-point functions (2 $\rightarrow $ 2  scattering kernels), {\em ad infinitum}. This fundamental and accurate description of QCD through its SDEs is not limited to its perturbative domain. However, in the weak coupling regime of QCD, this formalism reduces to the well-known order-by-order perturbative expansion of the $S$-matrix of the theory. Even more interestingly, in the non-perturbative domain of QCD, emergent phenomena of confinement and dynamical mass generation, inaccessible in perturbation theory, emanate naturally through realistic studies of the SDEs.  \\
\indent
Reliable extraction of physical observables in any non-perturbative formalism of QCD is cumbersome and prone to methods which are less systematic and robust than perturbation theory. In this article we employ the coupled SDE/BSE formalism.
Before attempting any solution within this approach, the infinite set of equations must be truncated by introducing mathematical model(s) of some suitable set of Green functions. That is why modelling remains a crucial component in hadron physics. However, despite this challenge, remarkable progress has been made and QCD akin truncations offer increasingly precise solutions,~\cite{Roberts:1994dr, Alkofer:2000wg, Maris:2003vk, Fischer:2006ub, Bashir:2012fs,  Raya:2024ejx}. 
Fortunately, our understanding of the intricate interplay between the quark propagator and the meson’s BS Amplitude (BSA)~\cite{Binosi:2016rxz} enables us to build their models that are amicable enough to allow for algebraic manipulations while still producing reliable predictions for the physical observables. \\
\indent
In this article, we construct a {\em QCD} kindred {\em algebraic model} (QAM) for the pion. Most algebraic models to date 
shuffle momentum dependence 
of the quark propagator and that of the BSA so as to 
simplify algebraic manipulation to compute hadronic observables. We present the construction of the
QAM which embodies the following 
key properties\,:
\begin{itemize}

\item The quark propagator incorporates its confining features. It is expanded in terms of complex conjugate poles which are present on the time-like half of the complex momentum plane. The existence and location of these poles resonates with more elaborate studies of the QCD SDEs.

\item  The momentum dependence of the quark propagator in our model is akin to the QCD kindred truncations of SDEs. The mass function saturates in the infrared while falls-off as $1/p^2$ in the ultraviolet  which agrees with its perturbative behaviour up to QCD logarithms in the chiral limit. 

\item  We put forward a simple QAM for the leading pion BSA which again encodes its low and large momentum behaviour.

\item The model guarantees that the axial vector Ward identity is observed, thus ensuring the Goldberger-Treiman relation connecting the quark propagator 
with the BSA is satisfied. 

\item To study the pion with electromagnetic probes, we also require the quark-photon vertex (QPV) for which we add to the usual $\gamma_{\mu}$ part a transverse vertex containing anomalous magnetic moment (AMM) distribution  which is known 
to affect the infrared limits of form factors.

\item The QAM in its totality ensures that the {\em Abelian} anomaly is correctly reproduced at $Q^2=0$.

\item 

We explicitly show that the large $Q^2$ power law for the TFF is correctly reproduced in accordance with the expectations of asymptotic QCD up to the logarithms which stem from the running of the strong interaction coupling. 

\item Finally, our constructed QAM takes into account the Efremov-Radyushkin-Brodsky-Lepage (ERBL) evolution equations for the pion distribution amplitude (DA).

\end{itemize}

Note that the precise shape of the pion DA to be determined is an intriguing topic that has been studied using numerous approaches, {\rm e.g.}, the QCD sum rules, the lattice QCD, the SDEs, etc. \cite{Brodsky:2007hb, chang2013imaging, RQCD:2019osh, cui2020kaon, Stefanis:2020rnd}. DAs play an essential role in describing various hard exclusive processes of QCD,~\cite{lepage1980exclusive, Chernyak:1983ej, Brodsky:1989pv}, such as the electromagnetic (EFFs) and transition form factors (TFFs)~\cite{lepage1979exclusive, lepage1980exclusive}, diﬀractive vector-meson production \cite{Forshaw:2012im, Gao:2014bca} and also in the study of \textit{CP}-violation via the nonleptonic decays of heavy-light mesons~\cite{El-Bennich:2009gqk, Shi:2015esa}, etc. 
%The pion TFF is highly sensitive to the DA.
We then employ our model to compute the 
pion TFF to two photons ($\gamma \gamma^*$).  \\
\indent
Chiral anomaly fixes the value of the pion TFF at the origin $Q^2=0$, 
while its behavior at large value of momentum  transfer $Q^2$ has been predicted through asymptotic QCD. Therefore, it is a particularly important transition 
which studies the internal structure of the neutral pion both in the infrared and ultraviolet domain of QCD through one single physical observable. 
Till 1998, the CELLO~\cite{cello1991measurement} and CLEO~\cite{gronberg1998measurements} data were available for the pion TFF in the space-like region $Q^2<7 \, \GeV^2$. The~\textit{BABAR} collaboration~\cite{aubert2009measurement} released data on the pion TFF for the space-like region $Q^2\in [4,40]$ around the end of 2009. This data initiated a contentious debate by suggesting the TFF's unexpected scaling violation through a power law instead of the usual much milder QCD logarithms. Their measured results show a rapid growth of $Q^2 G^{\gamma^{*}\pi^0\gamma}(Q^2)$ for the region $Q^2 > 15$, which contradicts the well-known asymptotic prediction that $Q^2 G^{\gamma^{*}\pi^0\gamma}(Q^2) \rightarrow \text{constant}$ for large $Q^2$ up to QCD logarithms. 
It led several works, which immediately followed, to suggest that the \textit{BABAR} data might not be an accurate measure of the pion  TFF,~\cite{Roberts:2010rn,Brodsky:2011yv,Bakulev:2011rp,El-Bennich:2012mkr}.
In support of these predictions, the subsequent Belle collaboration measurements in 2012 for the same energy region do not reconcile well with the \textit{BABAR} data~\cite{Belle:2012wwz}. Their data clearly seems to indicate that $Q^2 G^{\gamma^{*}\pi^0\gamma}(Q^2) \rightarrow \text{constant}$ for large $Q^2$. This was confirmed later by the QCD akin SDE computations~\cite{Raya:2015gva} and the recent data driven calculation~\cite{Higuera-Angulo:2024oui}. Hopefully, at the experimental level, the Belle II at SuperKEKB experiment~\cite{Belle-II:2018jsg} with less uncertainty at large $Q^2$ would be helpful to clarify the above experimental discrepancy. The aim of this work is to reexamine these varying claims using an updated investigation of the pion TFF with a QAM of the quark propagator, BSA and DA.

It might be worth recalling the importance and impact of the pion TFF to
two photons in the precision studies of the celebrated Standard Model of particle physics.
The hadron light by light contribution
coming from the neutral pion transition to
two photons is the most dominant to the
anomalous magnetic moment of the muon,~\cite{Eichmann:2019tjk,Raya:2019dnh}. Therefore, an accurate prediction of this observable is important in the precision studies of the standard model,~\cite{Aoyama:2020ynm}. 

\par
The article is organized as follows: in Sec.~\ref{sec:formalism}, we introduce the QAM of the quark propagator and the leading pion  BSA. We also present in detail the QPV with anomalous magnetic moment term (AMM) transverse to the photon momentum. In Sec.~\ref{section-PDA}, we discuss the pion DA and its QCD evolution. 
In Sec.~\ref{sec::ano}, we study, compute and report the $\gamma^{*}\pi^0\gamma$ TFF: in particular its connection with chiral anomaly, its asymptotic behavior in QCD and comparison with experimental data. We also explain the effect of the AMM term on pion TFF and the {\em Abelian} anomaly. Finally, in Sec.~\ref{sec::con}, we present our conclusions and final remarks.

%============================================================
\section{FORMALISM}\label{sec:formalism}
The BSE~\cite{salpeter1951relativistic, gell1951bound} provides the full relativistic description of meson bound states which appear as poles for specific values of invariant masses in BSA and these masses depend on the quantum numbers of the given meson state.
The solution of the BSE
\begin{eqnarray}\label{eq::bse}
[\Gamma_{M}(p;P)]_{tu}&=& {\int}\frac{d^{4}k}{(2\pi)^{4}}K_{tu}^{rs}(p,k;P) \nn
&\times& \Big(S_{a}(k_{+})\Gamma_{M}(k;P)S_{b}(k_{-})\Big)_{rs}\,,
\end{eqnarray}
provides the BSA, 1-particle irreducible quark-meson vertex $\Gamma_{M}^{a b}(p;P)$, where $a$ and $b$ represent the flavor of the quark and the antiquark respectively, and $r$, $s$, $t$, $u$ collectively stand for the color and Dirac indices. Here $M$ only specifies the type of the meson. The momentum conservation implies that $p_{\pm}=p\pm\eta_{\pm}P$ and $k_{\pm}=k\pm\eta_{\pm}P$ which satisfies the condition $\eta_{+}+\eta_{-}=1$. The renormalized, amputated quark-antiquark kernel $K$ is irreducible with respect to cutting pair of quark-antiquark lines. The kernel $K$ is the physical input to BSE along with the quark propagator and we must specify it in order to solve the BSE for the meson mass and the BSA. For a comprehensive review of the SDE-BSE formalism and its applications to hadron physics, see Refs.~\cite{roberts1994dyson, alkofer2001infrared, bashir2012collective, aznauryan2013studies}.\par
The a-flavor dressed quark propagator $S_a$ is obtained as the solution of the quark SDE~\cite{alkofer2001infrared, maris2003dyson, holl2006hadron, roberts2007aspects, sultan2021effect}
\begin{equation}\label{eqn::Ngap}
S_a^{-1}(p) \hspace{-1mm} = \hspace{-1mm} (i \slashed{p} + m_a) + \frac{4g^{2}}{3} \hspace{-2mm} \int \hspace{-2mm}\frac{d^{4}k}{(2\pi)^{4}} \hspace{-0.5mm} D_{\mu\nu}(p-k) \gamma_{\mu} S_a(k) \Gamma_{\nu}(k,p).
\end{equation}
Here $m_a$ is the $a$-flavor current quark mass and $g$ is the strong interaction coupling constant appearing in the Lagrangian. The rest of the ingredients of Eq.\,\eqref{eqn::Ngap} are defined as usual: the symbols $D_{\mu\nu}$ and $\Gamma_\nu$, respectively, are the fully dressed gluon propagator and the quark-photon vertex (QPV), each of which satisfy their own SDE. The eventual goal is to obtain the solution of the BSE by solving it in a coupled form along with the SDEs of the gluon propagator and the QPV.  This is a prohibitively tough problem to solve and such a study is beyond the scope of this work. Despite this impediment, many studies have been carried out which unveil the non-perturbative structure of these additional functions through QCD symmetries and/or directly using their corresponding SDEs~\cite{alkofer2004analytic, alkofer2005analytic, alkofer2009quark, fischer2003non}. Lattice formulation of QCD also provides useful insight~\cite{cucchieri2007s, skullerud2003nonperturbative, skullerud2002quark, skullerud2004quark}. Despite the fact that sophisticated, robust and reliable solutions for the propagators and BSA can be obtained from the corresponding SDEs and BSEs, less complicated models can be constructed which capture the  essence of infrared aspects of strong interactions, constrained by the requirements of asymptotic QCD~\cite{chang2014basic, albino2022pseudoscalar, sultan2021effect}. Our effort includes realistic and explicit  momentum dependence of the quark propagator, implements confinement and provides simple yet physically sensible description of the pion BSA as discussed in the 
subsections below.

\par

\subsection{Constructing the model}\label{sec:modeling}

A simple but efficacious model 
of the quark propagtor, the BSA and the QPV can be constructed to algebraically compute electromagnetically probed physical observables related to the pion. \\

\noindent
{\underline {\bf The Quark propagator}:} A simple but physically sensible {\em ansatz} for the momentum dependent quark propagator can be written as follows:
\begin{equation}\label{eqn::qp}
    S^{-1}(k)=i \gamma\cdot k + B(k^2)\,,
\end{equation}
with the running quark mass function~\cite{chang2016perspective}
\begin{equation}
\label{eqn::qpb}
    B(k^2) = \frac{y M^3}{k^2 +y M^2}\,,
\end{equation}
which saturates in the infrared to a constituent mass $B(0)=M$ and drops off as $1/p^2$ in the ultraviolet as dictated by perturbation theory save the logarithmic corrections. Here $y$ is a parameter which controls the width of the mass function. In order to mimic confinement efficaciously, the {\em ansatz} in~\Eqn{eqn::qpb} can be extended into the complex plane by requiring it to be expanded into a sum of a few complex conjugate pole representations\,:
%\begin{equation}\label{eqn::aqp}
%    S(k)=i \gamma.k \sum_{j}^{3} \frac{\alpha_j}{k^2+\gamma_j %M^2}+ \sum_{j}^{3} \frac{\beta_j M}{k^2+\gamma_j M^2},
%\end{equation}
\begin{equation}\label{eqn::aqp}
    S(k)=\sum_{j=1}^{3} \frac{-i\gamma\cdot k \alpha_j+\beta_{j} M}{k^2+\gamma_j M^2} \equiv -i\gamma\cdot k \sigma_A(k^2) + \sigma_B(k^2) \,.
\end{equation}
Comparing Eqs.~(\ref{eqn::qp}, \ref{eqn::qpb}, \ref{eqn::aqp}) and allowing for the complex poles leads to the %following 
identification of $\gamma_j$ and $\alpha_j$, $\beta_j$\,. 
%\begin{align}\label{eq::bc}
%\alpha_j &= \{0.439+0.993 i, 0.121, 0.0.439-0.993 i\}\,,    \nonumber \\
%\beta_j &= \{0.159+1.4896 i, -0.318, 0.159-1.4896\}\,,    \nonumber \\
%\gamma_j & = \{1.549+1.106 i, 6.903, 1.549-1.106 i\}\,.\nonumber
%\end{align}
It has long been known that the usually employed Rainbow-Ladder (RL) truncation to the quark propagator SDE does produce complex conjugate singularities on the time-like half of the complex  momentum plane which can be linked to confinement. Our present model for the quark propagator mimics these singularities which would play an essential role in this work. Such representation also allows us to directly calculate the form factors for large momentum regions~\cite{Chang:2013nia}. \\ 
%Moreover, we fix $M=0.506$ GeV
%$M=0.505521 GeV$ to produce pion decay constant as $0.092GeV$.\\

\noindent
{\underline {\bf Bethe-Salpeter amplitude}:} 
We propose a simple momentum dependent BSA model for the pion without considering the so-called spectral density function\,:
\begin{equation}
\Gamma(k;P)=i\gamma_5 \frac{1}{f_\pi} 
\frac{y M^3}{k^2+ y M^2}\,,
\end{equation}
where $k$ is the relative $\bar{q}q$ momentum and $f_\pi$ is the leptonic decay constant of pion.
%We follow Ref.~\cite{chang2013imaging} in order to write pion's momentum dependent BSA as 
%\begin{equation}\label{eqn::bsa}
%    \Gamma(k;P) = i\gamma_5 \frac{M}{f_\pi} \int_{-1}^{1} \rho(z) \frac{M^2}{(k+z\frac{P}{2} )^2+ M^2} \;\mathrm{d}z \,,
%\end{equation}
%where $P^2=-m_\pi^2$ and  $f_\pi=0.092GeV$. The $\rho(z)$ is a SDF whose precise shape determines the pointwise behavior of the meson's BSA.
At $P = 0$, {\em i.e.}, in the soft pion limit, we recuperate the exact Goldberger-Treiman relation\,:
\begin{equation}\label{eqn::bsap}
    \Gamma(k;P=0) = i\gamma_5 \frac{B(k^2)}{f_\pi} \,,
\end{equation}
where $B(k^2)$ is the Lorentz scalar mass function appearing in the quark inverse propagator, Eq.~(\ref{eqn::qpb}), thus already connecting the BSA of the pion with the quark propagator.  \\

\noindent
{\underline {\bf Quark-photon vertex}:} 
It has been known for a long time that the longitudinal Ball-Chiu (BC) vertex {\em ansatz}~\cite{Ball:1980ay} could  ensure chiral anomaly by construction~\cite{Bando:1993qy,Maris:1998hc}. Although we could naturally incorporate the BC vertex into our model but it has also long been established that without the appropriate transverse pieces of the vertex~\cite{Albino:2018ncl,Bashir:2011dp}, BC vertex is not entirely reliable. As a result,
it is noticed that in the intermediate photon momentum region, the calculated form factor is too hard~\cite{Maris:1999bh}.Therefore, we prefer our proposed QPV which includes a transverse AMM part. Our construction
consists of the bare vertex augmented by a part transverse 
to the photon momentum adapted from the {\em ansatz}~\cite{Dang:2023ysl,Qin:2013mta}~:
\begin{equation}
    \Gamma_{\mu}(p,q)=\gamma_\mu -\sigma_{\mu\nu}(p-q)_{\nu}\frac{B(p)-B(q)}{p^{2}-q^{2}}\mathcal{H}(Q^2) \,,
\end{equation}
where $\mathcal{H}(Q^2)=\eta (1-\exp[-Q^{2}/\hat{M}^2])/(Q^{2}/\hat{M}^2)$. The second term can be regarded as the dynamically generated non-perturbative analogue of the Pauli form factor for the AMM term,~\cite{Chang:2010hb,Bashir:2011dp}. Note that $\mathcal{H}(Q^2)$ is independent of the relative momentum of the quark and antiquark, and thus depends only on the photon momentum squared $Q^2$. The parameter $\hat{M}$ can be used to adjust the pion interaction radius.  
\par
\begin{figure}[hptb]
\includegraphics[width=8.6cm]{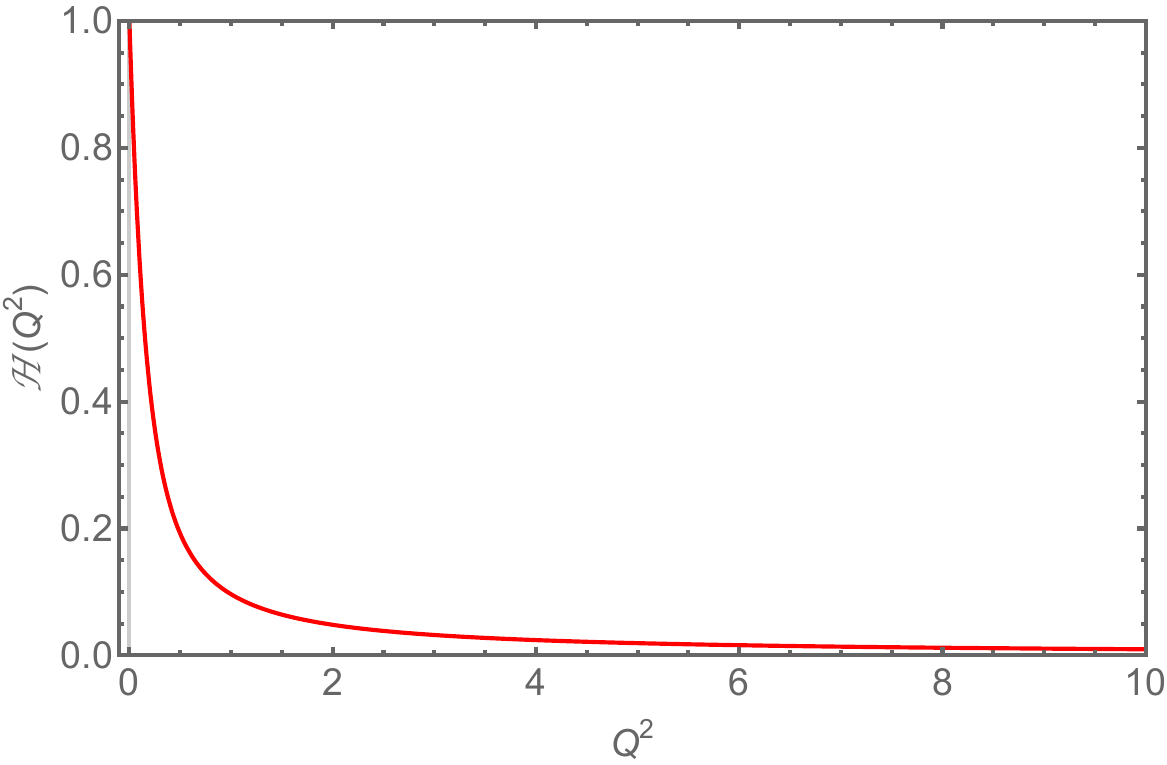}
\caption{The QPV dressing function $\mathcal{H}(Q^2)$ for the AMM contribution with $\hat{M}=0.3105$ GeV and $\eta=1$.  \label{HQ}}
\end{figure}

It is worth noting that the dressing function $\mathcal{H}(Q^2)$ of AMM term is proportional to strength parameter $\eta$. The AMM term disappears for $\eta=0$ limit and $\mathcal{H}(Q^2)\rightarrow 0$ as $Q^2\rightarrow\infty$. All these features of the $\mathcal{H}(Q^2)$ dressing function become more perceptible in Fig.~\ref{HQ}, which depicts its profile in a pertinent momentum range. Therefore, as we discuss in detail in Sec.~\ref{sec::ano}, the AMM term could indeed contribute to the chiral anomaly in the $\gamma^{*}\pi^0\gamma$ transition TFF quantitatively. %Some details of the QPV dressing function can be found in Refs. \cite{xing2021quark, dang2023chiral}.

\noindent
\section{Pion distribution amplitude and the model parameters} \label{section-PDA}

The pion TFF is sensitive to the shape of the pion DA. The precise behavior of the pion DA is an intriguing issue that has been addressed using numerous approaches, {\rm e.g.} the QCD sum rules, the Lattice QCD, the SDEs, and etc. \cite{chang2013imaging, brodsky2008light, mikhailov2016systematic, cui2020kaon, stefanis2020pion, gao2022pion}.
%
%{\bf Most of these favors the asymptotic form proposed by Lepage and %Brodsky \cite{lepage1980exclusive}...what this sentence means?}. 
%
%In this paper, We consider the following expression for the pion DA
%
Following the standard definition of the DA~\cite{Chang:2013pq} and the practical operation process~\cite{Xu:2018eii}, we can obtain the following analytical representation of the DA with the input of quark propagator and BSA, in the chiral limit,
\begin{equation}
\begin{aligned}
\varphi(x)&=\frac{3 y M^{2}}{2f_{\pi}^{2}\pi^{2}}\sum_{j1,j2}\frac{x\alpha_{j1}\beta_{j2}+(1-x)\alpha_{j2}\beta_{j1}}{2y-\gamma_{j1}-\gamma_{j2}}\\
    &\{\text{log}\left[\frac{2y(1-x)-(1-2x)\gamma_{j2}}{(1-x)\gamma_{j1}+x\gamma_{j2}}\right]\theta(x-\frac{1}{2})\\
    & +\text{log}\left[\frac{2yx+(1-2x)\gamma_{j1}}{(1-x)\gamma_{j1}+x\gamma_{j2}}\right]\theta(\frac{1}{2}-x).
\end{aligned}
\label{eq:PDA}
\end{equation}
We find that the proper choice of the parameter $y$ can produce different DAs at 2\,GeV. We choose two sets of parameters which reproduce the same value of $f_\pi$ and yield the second moment $<(2x-1)^2>=0.25(0.01)$ of the pion PDA based on an earlier SDE prediction~\cite{Roberts:2021nhw}. In Table~\ref{tab:parameters}, we present these best two sets of parameters at $2$\,GeV, the computed $f_\pi$ and the second moment of the PDA.  

\begin{figure}[hptb]
\includegraphics[width=8.6cm]{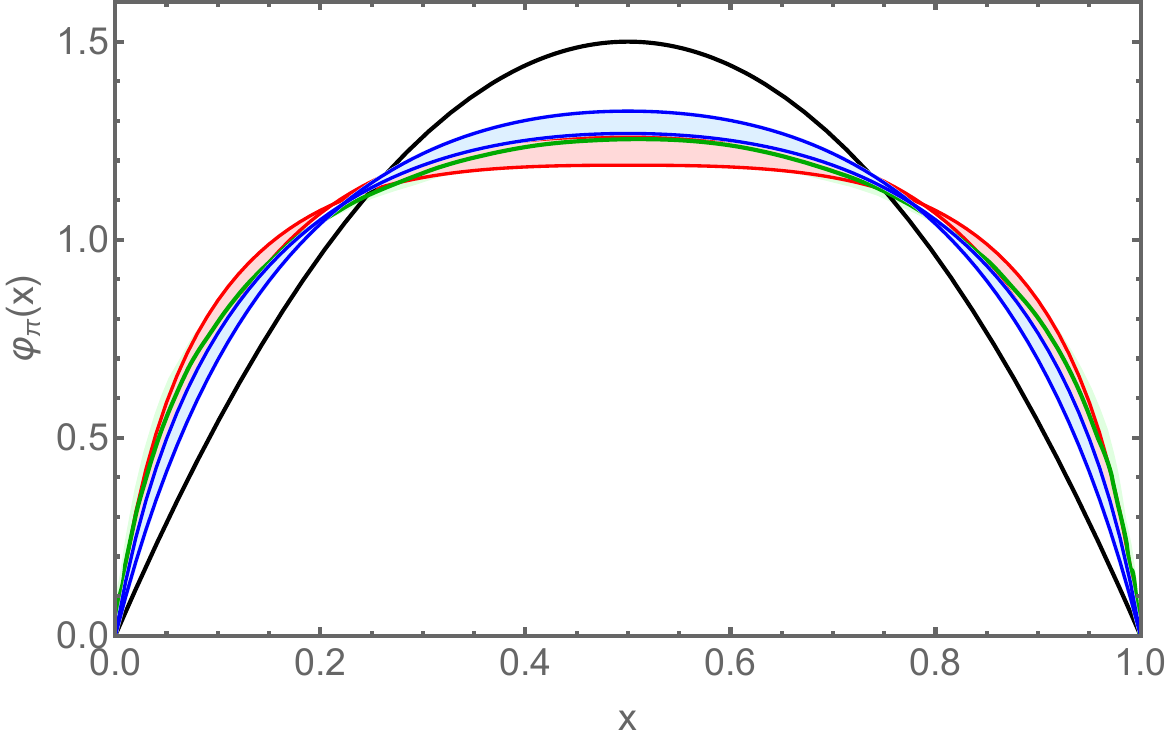}
\caption{DAs for $\pi$. The red band corresponds to our DA in Eq.~(\Ref{eq:PDA}). The blue band represents evolved DA at the scale of $\zeta^{2}=60{\GeV}^{2}$. The solid black curve corresponds to the asymptotic DA $ \phi_{asy}(x) = 6x(1-x)$. The green band represents the SDE prediction of the DA.
%corresponding to the second moment.
\label{PDA}}
\end{figure}

\begin{table}[htpb]
\centering
\caption{\label{tab:parameters} Parameter values at $2$\,GeV, and the corresponding computed pion decay constant and the second moment. }
\setlength{\tabcolsep}{2mm}{
\begin{tabular}{cccc}
\hline\hline
$y (\GeV)$ &$M (\GeV)$  & $f_{\pi} (\GeV)$ &$<(2x-1)^2>$ \\
\hline
7 & 0.3105 & 0.0924 & 0.24 \\
17 & 0.2580 & 0.0924 & 0.26  \\
\hline\hline
\end{tabular}}
\end{table}
We depict the pion DAs in Fig.~\ref{PDA}, and compare with the asymptotic two-particle DA $\phi_{asy}(x) = 6x(1-x)$ and the SDE prediction. Owing to the dominance of dynamical chiral symmetry breaking (DCSB), pion DA is rather dilated, exhibits a significant deviation from $\phi_{asy}(x)$ and has good agreement with the SDE prediction~\cite{Roberts:2021nhw}. All the DAs are symmetrical in Fig. \ref{PDA}. Historically, in the constituent quark mass algebraic models, the spectral density is introduced in the BSA to produce broad DA. However, in the present work, we find that the broad DA can be produced without the specific choice or even recourse to any spectral density. We propose the simplest form of the leading pion BSA with its correct power law description. It is merely equivalent to taking the delta function to represent the spectral density in earlier models.
\par
The QCD evolution of the pion DA is speciﬁed through the Efremov-Radyushkin-Brodsky-Lepage (ERBL) equations, derived in the Refs.~\cite{lepage1979exclusive, brodsky1979exclusive, lepage1980exclusive}. It determines the logarithmic dependence of the DAs on $\zeta$. At the leading order, the QCD evolution equations for the DAs can be expressed in terms of Gegenbauer polynomials \cite{courtoy2010generalized}
\begin{equation}\label{eqn::DA}
    \varphi(x,\zeta) = 6 x(1-x) \left(1+\sum_{n=1}^{\infty} a_n^{\zeta}C_n^{3/2} (2x-1))\right),
\end{equation}
with $a_{n}^{\zeta}=a_n^{\zeta_{2}} (\alpha_{s}(\zeta)/\alpha_{s}(\zeta_{2}))^{-\gamma_{n}}$. Note that only even $n$ contributes, and the anomalous dimensions are
\begin{equation}\label{eqn::ad}
    \gamma_n = \frac{C_F}{\beta} \bigg(1 + 4 \sum_2^{n+1} \frac{1}{k} - \frac{2}{(n+1)(n+2)} \bigg) \,,
\end{equation}
where we have $C_F = \frac{Nc^2-1}{2 Nc}$, $\beta = \frac{11}{3}N_c-2/3N_f$. Moreover, $\alpha_{s}(\zeta)=\frac{2\pi}{\beta {\rm ln}[\zeta/\Lambda_{\text{QCD}}]}$, with $\Lambda_{\text{QCD}}=0.234$~\GeV \, being the QCD scale parameter.
Once the relevant coefficients $a_n$ are determined, the evolution equation can be solved. These coefficients can be calculated using the initial conditions that the model calculation provides. By using initial DA $\varphi(x,\zeta_{2}=2 \, \text{GeV})$ along with the Gegenbauer polynomials orthogonality relation,
\begin{equation}
\begin{aligned}
a_n^{\zeta_{2}} =\frac{2}{3} \frac{2n+3}{(n+1)(n+2)}  \int_{0}^{1} \;\mathrm{d}x C_n^{3/2} (2x-1) \varphi(x,\zeta_{2}) \,,
\end{aligned}
\label{eq:PDA}
\end{equation}
 one is able to ﬁnd the $a_n$ coefficients. All model calculations should be consistent with the theoretical asymptotic behavior of the DA after evolution.
\par
Clearly, our model at the scale of $2$~GeV only depends on the parameters $y$ and $M$. We now introduce the scale dependence of these parameters. The procedure is as follows: We employ leading-order QCD evolution Eq.~\eqref{eqn::DA} to obtain the evolved DA at a specific scale. Then we refit the parameters $y$ and $M$, keeping $f_{\pi}$ unchanged, to generate new DA that is equivalent to the evolved DA. We repeat this procedure at different scales and, ultimately, obtain the scale-dependent extrapolation of these parameters. The expressions for these models are as follows:
%Clearly, our model only depends on the parameters $y$ and $M$. Here, we will introduce the scale dependence of these parameters. The procedure is as follows: starting from $2$~GeV, we use the leading order ERBL program to evolve the DA (distribution amplitude) to a specific scale. Next, we re-determine the parameters $y$ and $M$ while keeping $f_{\pi}$ constant, generate a new DA, and match it with the evolved DA. Ultimately, we obtain the scale-dependent model parameters. The details are as follows:
%\begin{eqnarray}
%\hspace{-5mm} y(\zeta) = \frac{9.144 + 0.959 \zeta^{2}}{
%  1 + 0.215 \zeta^{2}}\,\label{eqn::EPara1}, \; M(\zeta)=\frac{0.301 + 0.022 \zeta^{2}}{
%  1 + 0.063 \zeta^{2}}\label{eqn::EPara2};
%\end{eqnarray}
%and
%\begin{eqnarray}
%\hspace{-5mm} y(\zeta) = \frac{27.080 + 2.233 \zeta^{2}}{
%  1 + 0.285 \zeta^{2}}\,\label{eqn::EPara3}, \; M(\zeta)=\frac{0.241 + 0.030 \zeta^{2}}{
%  1 + 0.100 \zeta^{2}}\label{eqn::EPara4};
%\end{eqnarray}
\begin{equation}
\begin{aligned}
y(\zeta) &= \frac{7 + 0.722 (-4+ \zeta^{2}) + 0.003 (-4+ \zeta^{2})^2 }{
 1 + 0.151 (-4+ \zeta^{2}) + 0.0007 (-4+ \zeta^{2})^2}\,,\label{eqn::EPara1}\\
M(\zeta)&=\frac{0.3105 + 0.192 (-4+ \zeta^{2}) + 0.004 (-4+ \zeta^{2})^2 }{
 1 + 0.596 (-4+ \zeta^{2}) + 0.010 (-4+ \zeta^{2})^2};
\end{aligned}
\end{equation}
and
\begin{equation}
\begin{aligned}
y(\zeta) &= \frac{17 + 1.644 (-4+ \zeta^{2}) + 0.006 (-4+ \zeta^{2})^2 }{
 1 + 0.181 (-4+ \zeta^{2}) + 0.001 (-4+ \zeta^{2})^2}\,\label{eqn::EPara3}, \; \\M(\zeta)&=\frac{0.285 + 0.273 (-4+ \zeta^{2}) + 0.006 (-4+ \zeta^{2})^2 }{
 1 + 1.009 (-4+ \zeta^{2}) + 0.018 (-4+ \zeta^{2})^2};
\end{aligned}
\end{equation}
for the respective parameter sets. After the evolution, the DA takes similar functional form as Eq.~\ref{eq:PDA}, but with explicit scale dependence, $\varphi(x,\zeta)$.
Fig.~\ref{PDA} depicts the slow evolution of the DA towards the asymptotic limit through a light blue band. 

%At $\xi=2$~GeV, $y,M$ goes to Table-I.

%============================================================
\section{TRANSITION FORM FACTOR: $\pmb{\gamma^{*}\pi^0\gamma}$}\label{sec::ano}
A single scalar function is required to fully describe the  amplitude for the transition $\gamma^{*} \pi^0 \gamma $\,:
\begin{equation}\label{eqn::transitiontff}
    T_{\mu\nu}(k_1,k_2)=\frac{e^2}{4\pi^2f_\pi}\epsilon_{\mu\nu \alpha \beta} k_{1 \alpha} k_{2 \alpha} G(k_1^2,k_1 \cdot k_2,k_2^2),
\end{equation}
where  $k_1$ and $k_2$ are the photon momenta. The general impulse approximation for the transition amplitude of $\gamma^{*}\gamma \rightarrow\pi^0$ can be expressed as \cite{maris2002electromagnetic, raya2016structure, dang2023chiral}
\begin{eqnarray}\label{eqn::transition}
    T_{\mu\nu}(k_1,k_2)&=&\frac{N_c}{3}tr\int_qi\Gamma_{\nu}(k_2)S(q-k_2)\nn
    &\times&\Gamma_{\pi}(P)S(q+k_1)i\Gamma_{\mu}(k_1)S(q)\,,
\end{eqnarray}
where $P=-(k_1+k_2)$ is the total momentum of the pion  such that $P^2=-m_{\pi}^2$. The kinematic constraints are
\begin{equation}\label{kcon}
    k_1^2=Q^2,\ k_2^2=0,\ k_1\cdot k_2=-(Q^2+m_\pi^2)/2 \,.
\end{equation}
\noindent
By incorporating these kinematic constraints, the pion TFF is defined as
\begin{eqnarray}\label{eqn::tff}
   G^{\gamma^{*}\pi^0\gamma}(Q^2)=2 \, G(Q^2,0,0) \,,
\end{eqnarray}
where the factor 2 appears in order to account for the two possible orderings of the photons. The other elements in Eq.~\eqref{eqn::transition} are the quark propagator, pion BSA and the dressed QPV proposed in the previous sections. It is in principle straightforward to compute this transition. The isospin symmetry approximation, which assumes that up ($u$) and down ($d$) quarks have identical masses, is used.\par
We ﬁrst consider the chiral limit and $Q^2=0$ which corresponds to calculating the chiral anomaly, \textit{i.e.}, $G(0,0,0)$ using Eq.~\ref{eqn::transition}. The first thing to notice is that the use of the bare vertex alone, \textit{i.e.}, $\eta=0$, fails to reproduce the chiral anomaly. Introducing a more elaborate QPV, it is gathered that the AMM term does contributes to the chiral anomaly. To quantify this statement, let us consider $\eta=0$ limit and calculate Eq.~\ref{eqn::transition}. In this limit, $G^{\eta=0}(0,0,0)\neq \frac{1}{2}$. Thus the value specified by the chiral anomaly is not reproduced. Therefore, we fix the value of strength parameter $\eta$ to acquire $G(0,0,0)= \frac{1}{2}$. Within our choice of models for the quark propagator, QPV, the BSA and the pion DA, we find that the chiral anomaly is faithfully reproduced, $G^{\eta=0.598(0.615)}(0,0,0)= \frac{1}{2}$, at $2 \, \GeV$ with both parameters sets in Table~\ref{tab:parameters} for slightly different values of $\eta =  0.598(0.615)$. %We also find the exact anomaly using the evolved parameters.
%\textcolor{red}{$G^{\eta=0.404461(-0.280357)}(0,0,0)= \frac{1}{2}$ with scale dependent parameters in Eqs.~(\ref{eqn::EPara1}, \ref{eqn::EPara2}, \ref{eqn::EPara3}, \ref{eqn::EPara4}). what that means?}

%%%%%%%%%%%%%%%%%%%%%%%%%%%%%%%%%%%%%%%%%%%%%
%%%%%%%%%%%%%%%%%%%%%%%%%%%%%%%%
\begin{figure}[hptb]
\includegraphics[width=8.6cm]{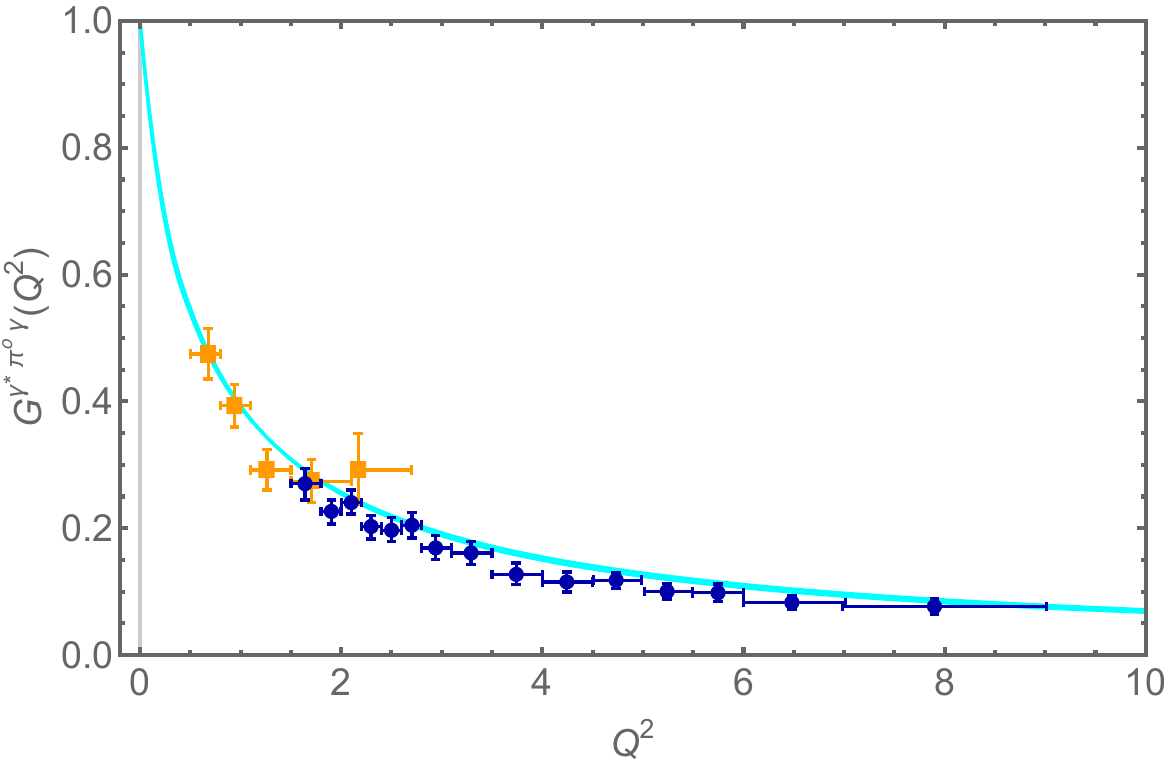}
\caption{$\gamma^{*}\pi^0The \gamma$ TFF. Curves:   \textit{brown dotted curve} corresponds to a monopole fit to the QCD-based result in Ref.~\cite{maris2002electromagnetic}, which agrees with the data reported in Refs.~\cite{cello1991measurement,gronberg1998measurements}, green squares and blue disks, respectively. Left panel- pion TFF results are represented with the light blue band computed with Eq.~\eqref{eqn::transitiontff} using parameters in the Table~\ref{tab:parameters}. %Right panel- The light blue band corresponds to our pion TFF results using the evolved parameters. 
Data: CELLO \cite{cello1991measurement}\textemdash Squares (orange); CLEO \cite{gronberg1998measurements}\textemdash Disks (blue).
\label{TFF}}
\end{figure}
%%%%%%%%%%%%%%%%%%%%%%%%%%%%%%%%%%%%%%%%%%%
%%%%%%%%%%%%%%%%%%%%%%%%%%%%%%%%%%%%%%%%%%%

The results obtained through our calculations of the $\gamma^{*}\pi^0\gamma$ TFF in both cases are depicted in Fig.~\ref{TFF}. It is clear that $G^{\gamma^{*}\pi^0\gamma}(Q^2)$ is a decreasing function of $Q^2$. It decreases rapidly for small values of $Q^2$, while it does so slowly for large values of $Q^2$.  It is also evident that our results agree with the experimental data for the complete range of $Q^2$, wherever the results are available and also with the results reported in Refs.~\cite{maris2002electromagnetic, chen2017two}.
%
%It is also evident that our results agree with the experimental data for all the range of $Q^2$, where results are available, and the results reported in Refs. \cite{maris2002electromagnetic, chen2017two}. Moreover, $G^{\gamma^{*}\pi^0\gamma}(Q^2)$ behaves like $1/Q^2$ for large $Q^2$. \textcolor{red}{Our numerical results suggest that the TFF is a nearly monopole and can be satisfactorily simulated by a vector meson dominance (VMD) formula $F(Q^2) = m_{\rho}^2/(m_{\rho}^2+Q^2)$. Lepage and Brodsky \cite{lepage1980exclusive} showed that TFF behaves like $Q^2 F(Q^2) \rightarrow 8 \pi^2 f_{\pi}^2 = 0.674 GeV^2$ in the asymptotic region while Fig. \ref{TFF} suggests slightly higher asymptotic mass scale.} Our results for pion TFF in both cases, with and without evolution of parameters, are approximately same.
%
Pion's interaction radius is defined as
\begin{equation}\label{eqn::radius}
    r_{\pi^0}^2 = -6\frac{d}{dQ^2} \ln g_{\pi\gamma\gamma}(Q^2)|_{Q^2=0} \,,
\end{equation}
where $g_{\pi\gamma\gamma}(Q^2)=G(Q^2,0,0)$.  
Our computed interaction radii are bounded within the range:
\begin{equation}
\label{eq:radii}
   r_{\pi^0}\in (0.63,0.67)\,;
\end{equation}
where 0.63 corresponds to the first set of parameters and 0.67 to the second set of parameters. Moreover, we use  $\hat{M}=0.42 \GeV$ in the damping function. The experimental value, $r_{\pi^0}=0.65\pm0.03\, \fm$ \cite{cello1991measurement}, lies within this range which is a positive result. On the other hand, in the limit $\eta=0$,
\begin{equation}
\label{eq:radii}
   r_{\pi^0}\in (0.53,0.60)\,;
\end{equation}
which is considerably less than the experimental estimate. This observation implies that the presence of the AMM term in the QPV is crucial.
%Our computed interaction-radius is $r_{\pi^0}=0.65\,\fm$, which is exactly same as the experimental value $r_{\pi^0}=0.65\pm0.03\, \fm$ \cite{cello1991measurement}. Whereas in the limit $\eta=0$, $r_{\pi^0}=0.57\,\fm$ which is more less than experimental estimate. This implies that the presence of the AMM term in QPV is crucial.
%\par
 For the sake of completeness, we also calculate the corresponding decay width, viz.,
\begin{equation}\label{eqn::decay}
\Gamma_{\pi^0\gamma\gamma}=\frac{g^2_{\pi\gamma\gamma}(Q^2)\alpha^2_{em} m^3_{\pi}}{16\pi^3 f^2_\pi}|_{Q^2=0} \,,
\end{equation}
where $\alpha_{\rm em}=1/137$ is the fine-structure constant. The decay width produced by Eq.~(\ref{eqn::decay}) is $\Gamma=7.72 \,\eV$ with both parameters sets, which is in good agreement with the experimental value $\Gamma=7.82\pm0.14\pm0.17\,\eV$\,,\cite{ParticleDataGroup:2022pth}. We observe that the decay width is substantially sensitive to the pion mass. Therefore, we use the experimental value of this mass, {\em i.e.}, $m_\pi = 134.9768\pm0.0005\, \MeV$ \cite{ParticleDataGroup:2022pth}.
\par
\begin{figure*}[hptb]
\includegraphics[width=8.6cm]{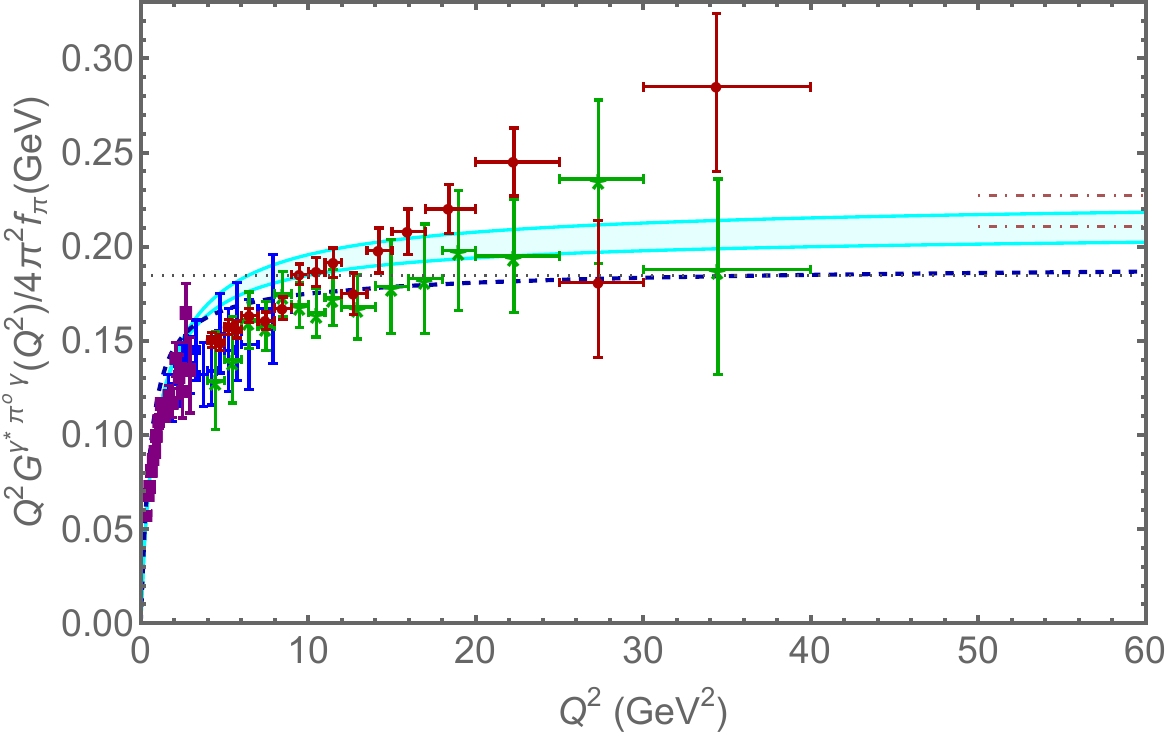}
\includegraphics[width=8.6cm]{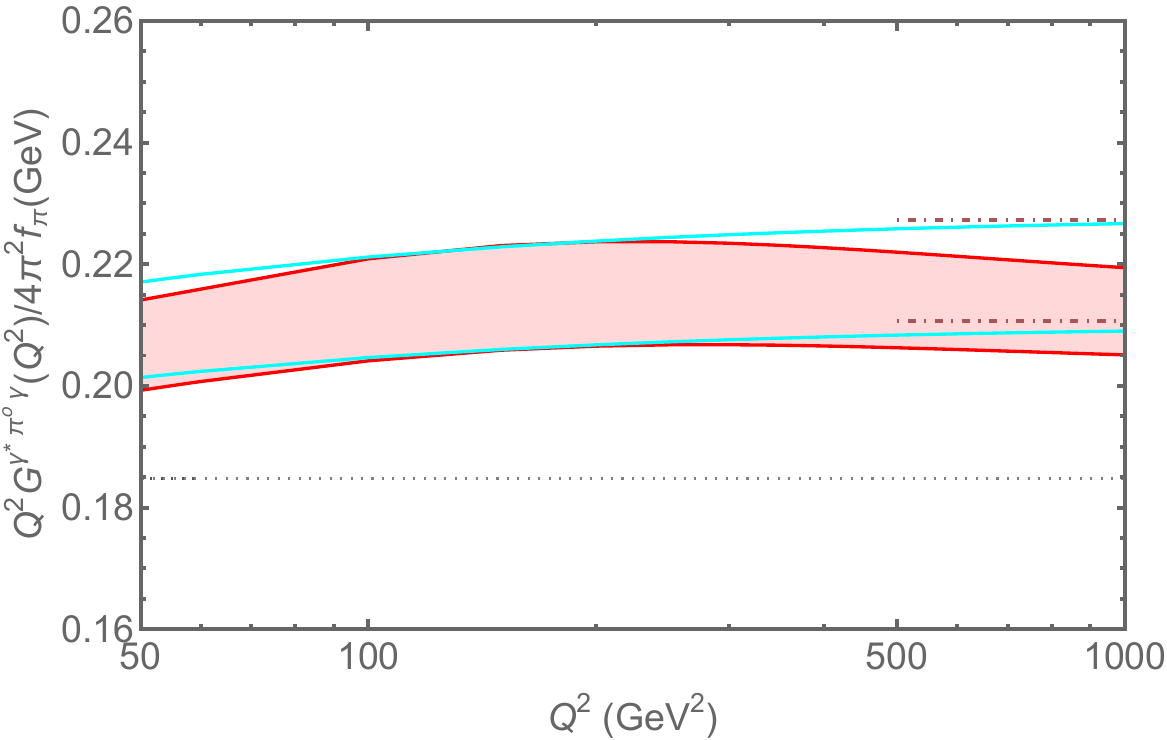}
\caption{$Q^2$-weighted $\gamma^{*}\pi^0\gamma$ TFF. Curves: \textit{blue dashed curve}\textemdash result obtained  from Eq.~\eqref{eqn::transitiontff} using asymptotic DA $\phi_{asy}(x)$%at $\xi_2$ without evolution
; \textit{gray dotted curve}\textemdash the limit with asymptotic PDA input; \textit{pink dot-dashed curve}\textemdash related to the limit with the present model DA at $2$~GeV input; Data: BESIII (preliminary) \cite{redmer2019measurements}\textemdash Squares (Purple); CLEO \cite{gronberg1998measurements}\textemdash Diamonds (blue); Belle \cite{gronberg1998measurements}\textemdash Stars (green); \textit{BABAR} \cite{aubert2009measurement}\textemdash Disks (red). Left panel- $Q^2 G^{\gamma^{*}\pi^0\gamma}(Q^2)$ results are plotted with the light blue band computed with Eq.~\eqref{eqn::transitiontff} using parameters in Table~\ref{tab:parameters}. Right panel- The light red band corresponds to our $Q^2 G^{\gamma^{*}\pi^0\gamma}(Q^2)$ results using the evolved parameters.
\label{TFFASY}}
%the $\pi^0$ asymptotic limit, $2 f_{\pi}$
\end{figure*}
\par
Our calculated $Q^2 \times $TFF is shown in Fig.~\ref{TFFASY}. We present two results, one where we use parameters at $2 \, \GeV$, left panel, and other with scale dependent parameters, right panel. Our prediction is indicated by the light blue band with parameters at $2 \, \GeV$, and light red band with scale dependent parameters. The asymptotic value of the pion TFF is $2 f_\pi = 0.185 \GeV$, plotted as the dotted (gray) curve. From the Figs.~\ref{TFF} 
and~\ref{TFFASY}, we can observe that our pion TFF is in agreement with perturbative QCD which predicts constant behavior of the pion $Q^2 \times $TFF at very large $Q^2$. Within the range of $Q^2 \leq 3 \,\GeV^2$, pion TFF agrees with the available BESIII (preliminary) data, and it has the same behavior as experimental determination, within the error bars, in $2 \,{\rm GeV}^2 \leq Q^2 \leq8 \,{\rm GeV}^2$ region. Fig. \ref{TFFASY} shows that the pion TFF has steeper $Q^2$ dependence in the range $Q^2 \leq 14 \,{\rm GeV}^2$ than the experimental measurements, and has good agreement with results from the Belle Collaboration beyond this range. The increase found by the \textit{BABAR} Collaboration for $Q^2 \geq 10 \,{\rm GeV}^2$ is not reproduced. In addition, there are theoretical studies speculating that the \textit{BABAR} data might not be the exact representation of the pion TFF~\cite{roberts2010abelian, Mikhailov:2009sa, Bakulev:2011rp, Wu:2011gf}. It is also shown in Ref.~\cite{Brodsky:2011yv} that the \textit{BABAR} data behavior at large $Q^2$ can not be explained with QCD calculations by using the asymptotic QCD, AdS/QCD, and Chernyak-Zhitnitsky models for the DA but can be supported by a flat  modeling of the pion DA model \cite{Brodsky:2011yv, Polyakov:2009je} which underestimate significantly the pion TFF at low $Q^2$. On the other hand, there are also phenomenological studies that support the \textit{BABAR} findings \cite{Wu:2010zc, Kroll:2010bf, RuizArriola:2010mrb, Agaev:2010aq, Gorchtein:2011vf}. It is worth noting that the $\pi^0$ TFF band calculated with evolved parameters touches the lower boundary of the TFF band calculated with parameters at $2\,\GeV$. It also lies within the error bars of the experimental data.

Finally, let's decipher the large $Q^2$ limit of the pion TFF 
within the proposed QAM. The leading order behavior is determined by using the bare QPV as input. By introducing Feynman parameterization and integrating over the momentum coordinate exactly, we can derive\,:
\begin{equation}\label{eqn::Feyn}
G(Q^{2})=4M^{4}y\sum_{j1,j2,j3}\int_{T} u_{1}^{2}u_{2} \, \frac{\mathcal{N}}{\mathcal{D}^{2}}\,,
\end{equation}
with ($\int_{T}=\int_{0}^{1}du_{1}\int_{0}^{1}du_{2}\int_{0}^{1}du_{3}$) \,,
and
\begin{eqnarray}
\mathcal{N}&=&(1+\beta)\alpha_{j1}\alpha_{j3}\beta_{j2}+(1-\alpha)\alpha_{j2}\alpha_{j3}\beta_{j1}\\
&&+(1-\alpha+\beta)\alpha_{j2}\alpha_{j1}\beta_{j3}\,,\nonumber\\
    \mathcal{D}&=&Q^{2} 2 u_{1}(1-u_{2})(1-\alpha)+2 y (1-u_{1})M^{2}\\ \nonumber&&+\gamma_{j1}M^{2}(1-2\alpha+u_{1})\\ \nonumber &&+\gamma_{j2}M^{2}(-1+2\alpha-u_{1}+2 u_{1}u_{2})\\ \nonumber&&+\gamma_{j3}M^{2}2 u_{1}(1-u_{2})\,,
\end{eqnarray}
where
\begin{eqnarray}
    &&\alpha=-\frac{1}{2}(-1+u_{1}(-1+2 u_{2}u_{3}))\,,\\
    &&\beta=-\frac{1}{2}(1+u_{1}+2 u_{1}u_{2}(-1+u_{3}))\,.
\end{eqnarray}
Using the change of variable $u_{3}=\frac{1-2\alpha+u1}{2 u_{1}u_{2}}$, one can convert $\int_T$ to
\begin{equation}
\left\{\int_{0}^{\frac{1}{2}}\int_{1-2\alpha}^{1}+\int_{\frac{1}{2}}^{1}\int_{-1+2\alpha}^{1}\right\}d\alpha du_{1}\int_{\frac{1-2\alpha+u_{1}}{2u_{1}}}^{1} \frac{du_{2}}{u_{1}u_{2}}\,.
\end{equation}
Now integrate over $u_{2}$ and retain the leading order term in powers of $Q^{2}$. Finally, integrating over $u_{1}$ by taking into account that $\sum\alpha_{j}=1$, yields\,:
\begin{equation}
    G(Q^{2}) \to \frac{4\pi^{2}f_{\pi}^{2}}{Q^{2}}\int_{0}^{1}d\alpha\frac{\varphi(\alpha)}{3(1-\alpha)} \,,
\end{equation}
where the DA $\varphi(\alpha)$ is defined through Eq.~\ref{eq:PDA}. 
We are thus able to explicitly derive the analytical expression of the TFF in the asymptotic limit within this model. It is\,:
%\begin{equation}\label{eqn::rhopda}
%   \frac{1}{4 \pi^2 f_\pi} Q^2 G^{\gamma^{*}\pi^0\gamma}%(Q^2 \rightarrow \infty) = 2 f_\pi \int_{0}^{1} %\;\mathrm{d}x \frac{\varphi(x,\zeta)}{3(1-x)} \,,
%\end{equation}
\begin{equation}
\exists \, Q_0 \gg  m_p  \; |  \;
\frac{1}{4 \pi^2 f_\pi} Q^2 G^{\gamma^{*}\pi^0\gamma}(Q^2) \stackrel{Q^2 > Q_0^2}{\approx} 2 f_\pi w(Q^2) \,,
\end{equation}
with $w(Q^{2})=\int_{0}^{1} \;\mathrm{d}x \frac{\varphi(x,Q^2)}{3(1-x)}$ and $m_p$ is the proton mass.
This behavior is fully consistent with the asymptotic QCD predication with the input DA\,: $\varphi(x,Q\to\infty)=6x(1-x)$. 
\par
If we choose the specific scale of $2$ GeV for the input quark propagator and the BSA for the computation of the DA, then we naturally expect a different Brodsky-Lepage (BL) limit which is clearly marked in Fig.~\ref{TFFASY}, dot-dashed curve, for both sets of parameters. Numerically, we obtain  the BL limit to be $2.272 f_\pi$ and $2.466f_\pi$ for parameters set 1 and 2, respectively. 
Fig.~\ref{TFFASY} shows that $Q^2\times$TFF increases with $Q^2$ and slightly overtakes the asymptotic limit of $2 f_\pi$. At very large value of $Q^2$,  $Q^2 \times$TFF approaches our computed $Q^2\to\infty$ limit from below.\par
Whereas with scale dependent parameters, the scaling violations are apparent, with the $Q^{2}$ dependence determined by the evolution of the DA. This arises a question what is the value of $Q_{0}$ for which scaling violations become apparent. The right panel of Fig.~\ref{TFFASY} clearly shows that $Q_{0}\sim 20\text{GeV}$ for our model at which this scale violation starts to occur. For $Q$ exceeding $Q_{0}$ the $Q^2\times$TFF deceases very slowly and we can expect it to approach $2f_{\pi}$ from the above as $Q\to\infty$.\par

%\begin{figure}[hptb]
%\includegraphics[width=8.6cm]{TFF.pdf}
%\caption{$\gamma^{*}\pi^0\gamma$ transition form factor. \textit{Red solid curve}\textemdash computation with Eq. \eqref{eqn::transitiontff}; \textit{Gray dashed curve}\textemdash VMD monopole result with mass scale $m_{\rho}^2 = 0.59 GeV^2$; \textit{Brown dotted curve}\textemdash corresponds to a monopole fit to the QCD-based result in Ref. \cite{maris2002electromagnetic}, which agrees with the data reported in Refs. \cite{cello1991measurement,gronberg1998measurements}, Green squares and Blue disks, respectively.
%\label{TFF}}
%\end{figure}

\par

%\textit{Brown dotted curve}\textemdash corresponds to a monopole fit to the QCD-based result in Ref. \cite{maris2002electromagnetic}, which agrees with the data reported in Refs. \cite{cello1991measurement,gronberg1998measurements};
%; \textit{Black dotted curve}\textemdash corresponds to a monopole fit to the QCD-based result in Ref. \cite{maris2002electromagnetic}, which agrees with the data reported in Refs. \cite{cello1991measurement, gronberg1998measurements}, Brown polygons and Orange disks, respectively.

\section{Conclusions}\label{sec::con}

In this work, we compute the pion TFF, $ G^{\gamma^{*}\pi^0\gamma}(Q^2)$, for one real and one virtual photon. We employ models of the quark propagator, pion BSA and the QPV instead of the solutions of the SDEs and BSEs of  QCD. For completeness, we also calculate the pion interaction radius and the corresponding decay width. We model the expected momentum dependence of the quark propagator with a QAM which captures the complex conjugate singularities on the complex momentum plane which can be related to confinement. We construct the BSA without a spectral density function  which encapsulates its low and large momentum behaviour. We then obtain the pion DA by using the proposed quark propagator and the BSA. We then evolve it according to the ERBL evolution equation from perturbative  QCD at hadronic scale to complete the computation of the pion TFF. Our calculations are based upon the bare as well as the dressed QPV, by adding the quark AMM term to the tree level vertex. While both {\em Ans$\ddot{a}$tze} of the QPV produce the same qualitative behavior of the pion TFF, the quark AMM term play an essential role to produce the correct chiral anomaly and the charge radius of the pion.
\par
Note that the tree level vertex fails to produce the chiral anomaly, which is associated with on-shell photon and chiral limit pion, \textit{i.e.}, $G(0,0,0)$. The AMM term serves as an effective term that contributes to normalize the TFF. The strength parameter $\eta$ can be tuned to produce chiral anomaly, \textit{i.e.} $G(0,0,0)=\frac{1}{2}$. The dressed QPV follows the asymptotic limit, \textit{i.e.} $ \Gamma_{\mu}(Q) \rightarrow \gamma_\mu$ as $Q^2 \rightarrow \infty$. It is clear from Fig.~\ref{TFF} that the AMM term contributes significantly to the QPV in the low $Q^2$ domain. Its contribution virtually disappears as $Q^2$ increases. As a result, the interaction radius and the decay width come out to be consistent with the empirical findings.
\par
We present pion TFF results using the parameters at $2 \, \GeV$ as well as using the evolved parameters. We find decent agreement with experimental data. However, we tend to disagree with the rapid increase in large $Q^2$ region suggested by the \textit{BABAR} collaboration; instead we have reasonable accord with the Belle data. The computation at any desired value of space-like $Q^2$ is an extra advantage of the QAM due to its simplicity. We also find that our numerical result for the pion TFF agrees well with BESIII (preliminary) data in the domain $Q^2\leq3 \, \GeV^2$. It is hoped that the upcoming, more precise results from the planned Belle II experiment would help explain the previous discrepancies between experimental observations and theoretical predictions \cite{Belle-II:2018jsg}.
Our DA model has good agreement with the earlier SDE prediction which lies within our proposed DA band. Due to the different scale, {\em i.e.}, $2 \, \GeV$, our BL limit is larger than the asymptotic prediction limit, $2f_\pi$. However, our numerically computed pion TFF is clearly in agreement with our analytically calculated BL limit with both the parameters sets, and at very large $Q^2$, it  approaches the BL limit from below. The evolution of DA provides the phenomena of scaling violation, starting from $Q_{0}\sim 20\text{GeV}$ and TFF approaches to $2f_{\pi}$ from above at $Q\to\infty$.
\par
In conclusion, we suggest a model of the pion BSA which is simple but offers a powerful tool for the analysis of the electromagnetic properties, and the momentum-dependent model of the quark propagator which captures the confining features through related complex conjugate singularities. We also propose a model for the PDA that has excellent agreement with the SDE predictions. The pion TFF results obtained with this compound QAM agree well with the available experimental data and obey our computed BL limit at $2$ GeV from the below which can be obtained analytically. Whereas with the evolution of DA, the pion TFF approaches the asymptotic limit, $2 f_\pi$, from the above.
Although the model is simple, without recourse to a spectral density function and an added advantage of the momentum dependent quark mass function with confinement mimicking complex conjugate singularities, it must be able to produce a large number of pion related observables, such as the pion EFF, geneneralized parton distributions and the parton distribution functions at the experimental scale. All this is for future.

%We also like to point out that the improvement in BSA like evolution with running scale would be necessary.

\hspace*{\fill}\ 
\begin{acknowledgments}
This work is financially supported by the National Natural Science Foundation of China (grant no. 12135007).
A.~B. wishes to acknowledge the {\em Coordinaci\'on de la Investigaci\'on Cient\'ifica} of the{\em Universidad Michoacana de San Nicol\'as de Hidalgo}, Morelia, Mexico,  grant no. 4.10, the {\em Consejo Nacional de Humanidades, Ciencias y Tecnolog\'ias}, Mexico, project CBF2023-2024-3544
as well as the Beatriz-Galindo support during his current scentific stay at the University of Huelva, Huelva, Spain.

\end{acknowledgments}
%\newpage

%\appendix\widetext
\bibliography{bibreferences}
\end{document}